# On the bubble-bubbleless ocean continuum and its meaning for the lidar equation: Lidar measurement of underwater bubble properties during storm conditions

D. Josset[1*], S. Cayula[1], B. Concannon[2], S. Sova[3], A. Weidemann[4]
*1 U.S. Naval Research Laboratory (NRL), Stennis Space Center (SSC), Ocean Sciences Division, 1009 Balch Blvd, 39529 Stennis Space Center, MS.*
*2 Naval Air Warfare Center – Aircraft Division (NAWCAD).*
*3 NRL, SRR, 1009 Balch Blvd, 39529 Stennis Space Center, MS.*
*4 NRLSSC retired.*
***damien.josset@nrlssc.navy.mil***

**Abstract:** This paper presents the NRL shipboard lidar and the first lidar dataset of underwater bubbles. The meaning of these lidar observations, the algorithms used and their current limitations are discussed. The derivation of the lidar multiple scattering regime is derived from the lidar observations and theory. The detection of the underwater bubble presence and their depth is straightforward to estimate from the depolarized laser return. This dataset strongly suggest that the whitecaps term in the lidar equation formalism needs to be revisited. Void fraction retrieval is possible and the algorithm is stable with a simple ocean backscatter lidar system. The accuracy of the void fraction retrieval will increase significantly with future developments.

## 1. Introduction

The breaking of surface waves injects air into the water column. It happens all over the ocean, even if most people are only familiar with the breaking of waves near the shore and the associated foamy, bubbly surface called surf. This entrainment of air forms bubble clouds underwater [1-3]. These entrained bubbles, in turn, change the optical and acoustic properties of the water column [4]. In addition to the sound speed change and acoustic transmission loss increase, the breaking generates acoustic noise [5].

In high-wind conditions, bubbles become a critical component of the air-sea gas exchange [6, 7], especially the uptake of carbon dioxide and oxygen [8]. In general, bubbles play a significant role in air-sea exchanges of mass, heat, energy, and momentum [9, 10].

Although it has been known for quite some time that the lidar return is sensitive to whitecaps and bubbles [11, 12], there are not many publications on this topic, and additional published studies discussing the lidar return of bubbles in the ocean would allow us to understand the added value of this instrument. The impact of whitecaps on the CALIPSO space lidar return is briefly shown and discussed in [13] and [14]. In the context of the fundamental lidar equation [15], we stressed the importance of more studies using lidar depolarization "at high wind speeds when bubbles are forming inside the water column." More recently, [16] showed and discussed case studies of the lidar return of bubbles created by ship wakes.

This paper presents:
- The Naval Research Laboratory (NRL) shipboard lidar and its calibration procedure.
- The first dataset of bubble profiles observed by shipboard lidar in high wind conditions.

- A demonstration that the depolarization of the bubble features comes from small-angle multiple scattering. For this reason, the vertical depth is measured accurately with the shipboard lidar.
- The fact that whitecaps and bubbles have a strong and unambiguous depolarization signature. It is used for a feature detection algorithm of ocean bubbles (a "bubble mask").
- The discussion of a necessary paradigm shift for the whitecap term embedded within the lidar equation for in-water laser light propagation.
- The lidar void fraction retrieval procedure and its accuracy.

Lidar can provide simultaneous vertical information on both the atmosphere and the ocean. As such, it can provide information on the bubble's vertical properties (bubble depth and void fraction) within the context of wave height and sea spray injection. Despite the shortcomings of the data due to sparse coverage and the inability to completely decouple and decorrelate the different geophysical contributions from the different scatterers, lidar use opens exciting possibilities for future studies of the air-sea interface.

## 2. Instrument Design
### 2.1. System Overview

The NRL shipboard lidar is one of the significant assets of the NRL Ocean Sciences division at the NASA Stennis Space Center. It measures the elastic backscattering of laser light at 532 nm. The primary data products are the ocean backscatter coefficient, total attenuation coefficient, and degree of linear polarization [19]. The lidar has been used on ship deployments continuously from 2013 to 2019 (East Sound in Washington state, Chesapeake Bay, Gulf of Maine, Atlantic Ocean, Lake Erie, and Gulf of Mexico). The system has been receiving regular modifications since 2013 to fit the needs of different projects.

A typical setup of the lidar, is to be on the bow of research vessels, and typically setup at an angle between 15 and 20 degrees, to limit the ocean surface backscatter intensity. It was mounted at an angle of 6.3 degrees on the R/V Sikuliaq because the anticipated adverse environmental conditions led us to design a much sturdier mount to protect the optics. The lidar, as set on the R/V Sikuliaq, is shown in Fig. 1. We built the mount in steel, based on 0.635 cm thick plates and 20.32 cm I-beams, secured to the bow with 2.54 cm diameter bolts. The power supply was mounted directly under the lidar against the hull, and the bow served as a protecting structure against the waves. The intensity of the ocean surface signal at the 6.3 degrees angle did not induce difficulty for the data analysis, because the rough ocean condition (ship roll and rough surface) led to a relatively low surface return, and because the 15-20 degree setup is conservative when the lidar system is properly designed.

We designed the instrument to be as compact as possible while ensuring enough structural robustness to enable deployment on a ship while underway. This compact and robust design allows the system to sample ocean properties, even in the harsh environment of the open ocean. The soundness of the design and the setup was demonstrated without ambiguity during the 18 days of deployment (12/05/2019 to 12/23/2019) in the Gulf of Alaska in high winds and storm conditions, with wave heights recorded up to 17 m. Fig. 1 illustrates one of the many wave events experienced by the lidar. The boat camera takes pictures regularly, but their frequency (12 per hour) is not sufficient to capture a representative sample of wave events. Therefore, most of the waves event experienced by the lidar are documented only through the lidar data. The crew reported that the lidar was under 2 m of water three times due to large waves reaching the bow on the night of 12/11/2019. It did not show any degradation of capabilities after this or any other wave events it experienced.

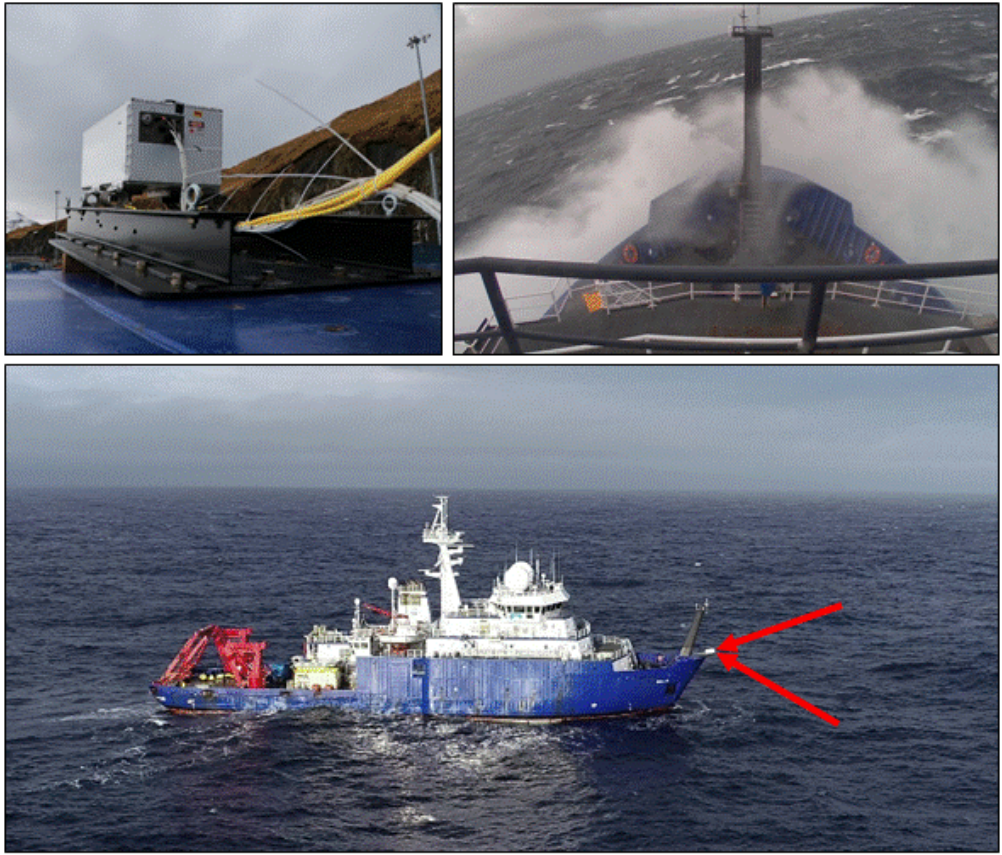

Fig. 1. Top left and bottom: The NRL shipboard lidar mounted on the bow of the R/V Sikuliaq. The red arrows show the position of the lidar (the small white dot on the bow). Top right: Picture of one of the waves that reached the bow. This picture is representative of a relatively minor size wave. There are no photographs of the more significant wave events.

2.2. **Lidar system description**
The system comprises a laser transmitter, beam polarization optics, photoreceivers, data collection, and control hardware. The transmitted beam path is in Fig. 2, describing the key elements.

The receivers have a 140 mrad Field of View (FOV). The 1 ns pulse width of the laser combined with the 800 MHz digitization rate permits a vertical sampling of about 0.14 m underwater. In contrast, the 50 Hz sampling yields an along-track resolution of approximately 0.02 to 0.1 m. In order to provide sufficient overlap between the transmitted beam spread and detector FOV, the shipboard lidar is mounted at least 4.2 m above the sea surface.

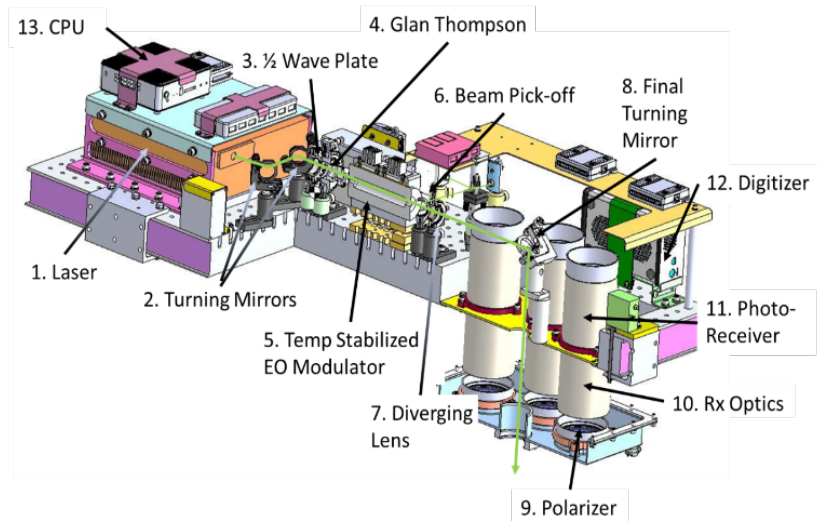

Fig. 2. Figure of the optical transmission path of the NRL shipboard lidar.

NAWCAD designed and built the system, and Welch Mechanicals, LLC built the watertight enclosure. Fig. 2 shows the optical transmission path of the NRL shipboard lidar in its original state.

In the original (2013) design, an Electro-Optic (EO) modulator could modify the polarization of the laser light from circular to linear. In this configuration, two receivers would measure the linear polarization states (co- and cross), and two receivers would measure the circular polarization states. In 2015, we removed the modulator to reduce the amount of electronic noise in the signal and optimize the system for oil research [17]. The laser polarization is now fixed and linear.

The transmission of the laser beam goes through an optical window situated in the center of up to six receiver units. Each receiver consists of a telescope, which focuses the lidar return signal to a photomultiplier tube (PMT). The electrical signal from the PMT connects to a digitizer channel. The receivers are identical except for a different polarizer or/an optical filter at the entrance aperture to allow detection of the polarization of interest (or wavelength for fluorescence). There are six receiver positions, and the acquisition software is written for up to 6 channels, making the system modular and reconfigurable. Up to four receivers and two digitizers (with two channels each) have been used so far during deployments. Two receivers measure the (linear) co-polarized backscattered light, one measures the (linear) cross-polarized return, and one is sensitive to the fluorescence of oil with a relatively wide (50 nm) bandpass filter centered at 575 nm.

We present the characteristics of our system in the following discussion. Key characteristics of the transmitter and receiver are in Table 1 and Table 2.

The data acquisition system provides remote control and diagnostics of the instrument, even if there is no safe access to the ship's bow. As a result, it can run 24/7, even without the possibility of manual adjustment or repair, even during several weeks of boat deployment in storm conditions. A master laptop computer sends all parameters (PMT gains, gate timing) and control commands to the lidar, and the lidar system box stays sealed.

A GPS/IMU unit collects attitude and position information for each laser shot, which goes into the lidar data stream in real time.

Due to the proximity to the surface, the signal-to-noise is typically higher than for airborne or spaceborne systems (3 to 5 order of magnitudes). In addition, a custom compression algorithm created and implemented by NAWCAD allows the system to go beyond the 11 Effective Number of Bytes of the digitizer and reach a dynamic range >14 Bytes.

However, even low signal-to-noise lidars will be sensitive enough to monitor the water body down to some depth, and a well-designed ocean lidar system has a sufficient vertical resolution and penetration depth to detect the features of interest. A critical difference between this lidar and most other operating systems is the high vertical resolution of the oceanic feature it can detect (14 cm underwater). This is a key characteristic of this system, that makes the dataset so unique.

Additionally, the low speed of the boat and relatively fast sampling rate create an almost stationary measurement where the lidar does not move. However, the ocean feature evolves under it as a function of time. From a scientific point of view, it is complementary to airborne and spaceborne lidar observations. These platforms travel so fast that they are better adapted to study the spatial scale of ocean features.

TABLE I. NRL SHIPBOARD LIDAR TRANSMITTER SPECIFICATIONS

| Wavelength | 532 nm |
|---|---|
| Pulse energy | 1 mJ |
| Repetition rate | 50 Hz |
| Ground spot spacing | 0.1 m (5 m/s) |
| | 0.02 m (1 m/s) |
| Beam divergence | 12 mrad (after beam expander) |
| Pulse width | 1 ns |

TABLE II. NRL SHIPBOARD LIDAR RECEIVER SPECIFICATIONS

| Telescope diameter | 5 cm (6 units) |
|---|---|
| Field of view | 140 mrad |
| Optical filter bandwidth | 1 nm |
| Detector quantum efficiency | >20% |
| Detector dark current | 1 nA |
| Digitizer sample rate | 800 MHz |
| Vertical sampling spacing | 0.14 m (underwater) |
| Digitizer resolution | 14 bits |

## 3. Field mission research objectives

The lidar was on the bow of the R/V Sikuliaq from 4th December 2019 to 23rd December 2019. The winter deployment in the Gulf of Alaska was in collaboration with the UNOLS cruise of the "Wave breaking and bubble dynamics" (Breaking Bubble), led by Principal Investigator (P.I.) J. Thomson, co-P.I. M. Derakhti and funded by the National Science Foundation (NSF). The project aimed to understand the turbulence beneath waves breaking at the ocean surface. The dynamics associated with bubble plumes generated during the breaking process are a particular focus. P.I. Thomson invited the NRL researchers to bring the lidar on the cruise. The NRL internal IMPACT 6.2 project supported the NRL researchers' participation in this work. This project aims to derive the vertical properties of bubble clouds with lidar technology and use this information to understand the ocean environment better.

## 4. Meteorological conditions and sea state

As shown in Fig. 3, during the 18 days at sea exercise, the R/V Sikuliaq experienced several storm conditions associated with the passage of low-pressure fronts below 1,000 hPa. The average wind speed value was 9.7 m/s (±4.8), with a minimum of 0.05 and a maximum of 26.3 m/s (with wind gusts up to 32.9 m/s). Wave heights ranged from 3 to 10 m, with extreme wave events in the area as recorded by the Swift buoys [18, 19] up to 17 m.

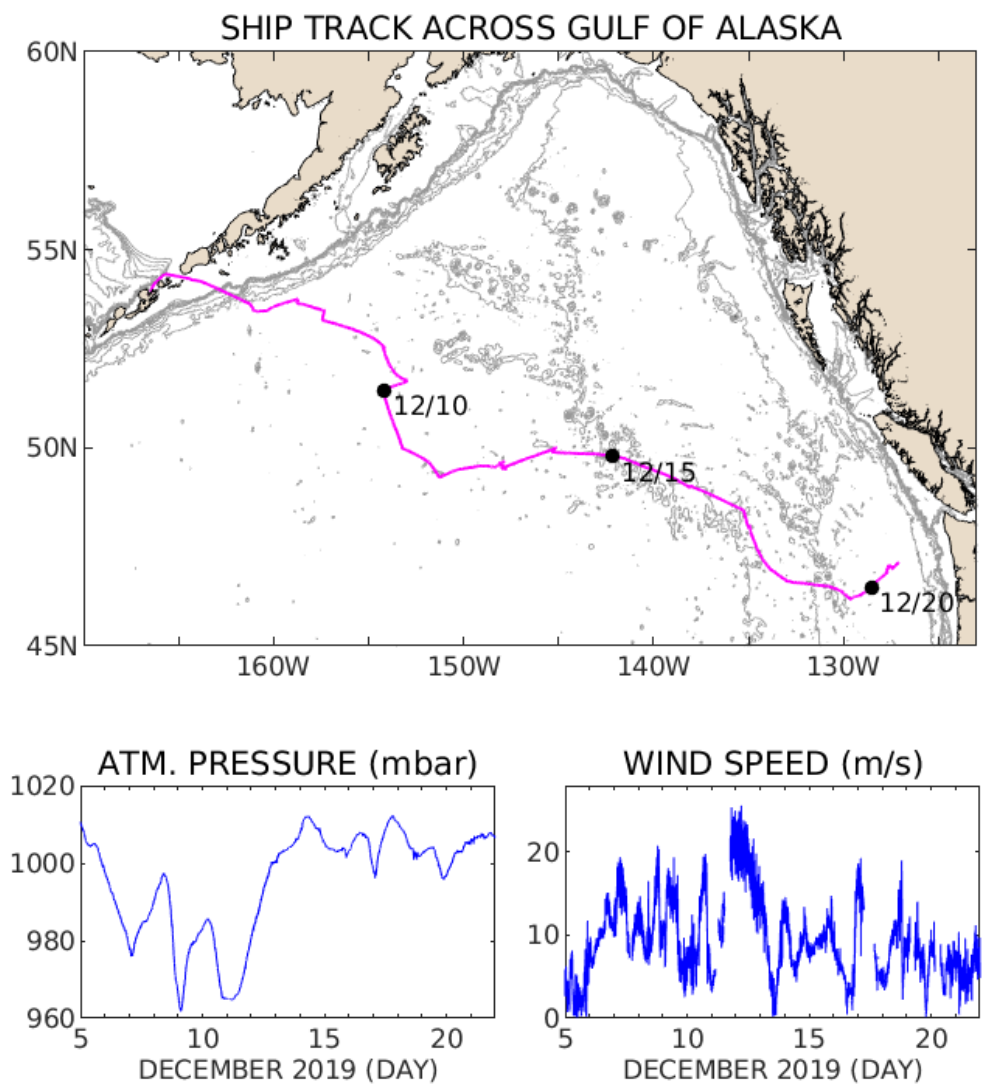

Fig. 3. Top: track of the R/V Sikuliaq across the Gulf of Alaska. Bottom left: pressure levels (m/s) during the travel. Bottom right: same as left but for wind speed (m/s).

The wind originated primarily from the West and South-West (average 216.08 ± 64.44 degrees). Water salinity, temperature, and chlorophyll-a content were relatively stable at 32.14

± 0.14 gram of salt per 1000 grams of water (or Practical Salinity Unit, p.s.u), 10.22 ± 1.53°C, and 1.94 ± 0.32 mg.m$^{-3}$, respectively.

Regarding shipboard conditions, the boat experienced regularly 20 to 30-degree roll as the multidirectional wave systems made it difficult to find a stable heading for the ship. The ship's roll for the data presented in this paper is in Fig. 4. The conditions were ideal for finding bubbles generated by breaking wave events due to the high wind speed (Fig. 3, bottom right). These conditions do not affect too much the lidar initial setup, and the off-nadir angle is 6.7 degree on average, very close from the initial 6.3 degree setup. On average, the lidar height is 9 m above the water surface. This determines the calibration altitude, as discussed in section 5.

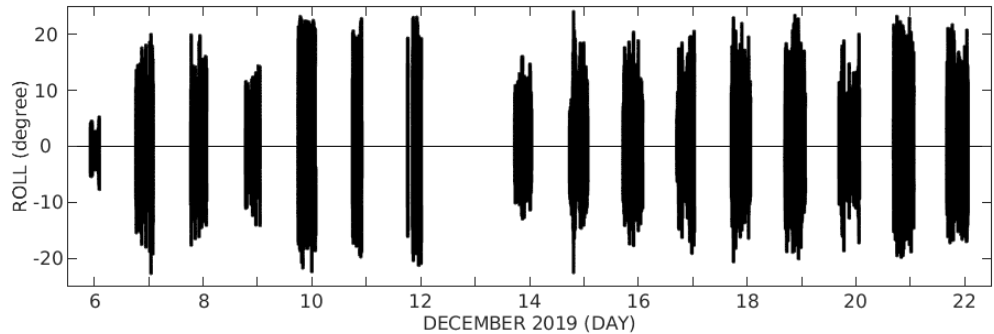

Fig. 4. Time series of the ship's roll as a function of time for the lidar data under analysis.

The lidar acquired data while the boat faced the wind and maintained a forward speed of around 0.5 to 1 m/s. Overall, we gathered data from the 6$^{th}$ to 22$^{nd}$ (system setup and initial tests on the 5$^{th}$, start to clean up and package on the 23$^{rd}$), which resulted in the collection of more than 113 hours of data at 50Hz (around 20M ocean profiles) that span different winds and wave conditions. This lidar dataset allows us to understand the statistical occurrence of bubbles in the ocean as measured by a lidar. It is the first published result of this kind.

## 5. Data calibration procedure

The dataset is analyzed and presented for every individual laser shots. The lidar records data in its own reference frame, and provides information as a function of the lidar distance. The boat and lidar GPS/IMU allow to correct for the lidar altitude and attitude variations. For clarity of the presentation, the shots are presented as vertical in the pictures, but as previously explained, they are on average taken at a 6.7 off-nadir angle. As a consequence, the bubble depth we discuss is biased low. The average bias would be easy to correct, but it would raise the question on why to not correct on a shot to shot basis. If such a correction is applied, the data would need to be presented in 3 dimensions, which is not how bubbles data are usually presented. It would be unique and extremely interesting to discuss the 3 dimensional structure of the bubble clouds, but it is not within the scope of this paper which focuses on the lidar observations and the algorithm.

We used both the atmospheric backscatter and the ocean surface as calibration targets. Interestingly, we found that using the ocean surface calibration return as a calibration target is not trivial for the shipboard lidar. In contrast, it works very well for space lidar [14]. The exact cause would require more investigation. However, one issue seems to be that the wave systems we experienced were loosely related to wind speed. The ship did not roll significantly less

(Fig.3 and Fig. 4) when the speed decreased from 20 m/s to 10 m/s. In addition, preliminary analysis (not shown) seems to indicate that the slope of the waves varied so much that the reference for the mean square slope of the waves would need to be adjusted (i.e., the ocean surface return changes significantly between the two sides of a large wave). We have never noticed this issue in the calibration of the CALIPSO lidar, probably because of the larger laser footprint and because waves are, in statistics, smaller than what we experienced during this cruise. This variability with the incidence angle is the reason why we did not use the ocean surface as the calibration reference for this paper.

Previous publications describe the general principle of the atmospheric backscatter calibration procedure based on Rayleigh scattering [20, 21]. For a shipboard lidar, the accuracy of this calibration will be much lower than, for example, a space or airborne lidar, which can rely on clear air in the upper atmosphere as a calibration target [22, 23]. When the lidar is at low altitude, the presence of aerosols in the atmospheric return cannot be overlooked, and the calibration procedure requires a specific methodology to limit the uncertainty due to aerosol contamination.

The air temperature and pressure are part of the standard measurements from the RV Sikuliaq. They are performed with a fan-aspirated MET4A Meteorological Measurement Systems by Paroscientific, Inc, mounted on the forward mast. The pressure accuracy is better than ±0.08 hPa, and the temperature accuracy is better than ±0.1°C. These measurements are used directly to determine the air density and the backscatter of air molecules. This is the calibration reference for the lidar signal. In order to take into account the movement of the boat and the ocean surface height changes, the calibration reference is the average lidar signal between 3.5 and 4.2 m above the ocean surface, without any extrapolation of the vertical profile shape of the atmospheric density (i.e. the procedure used in [24] was unnecessary). With an average 9 m distance from the water, this procedure is a compromise to have enough data for the aerosol filtering procedure, far enough from the lidar but not too close to the ocean surface. In order to limit the influence of heavy sea spray events on the statistics, we make the calibration on profiles with an atmospheric backscatter coefficient equal to the median value for the whole file. Although this does not correspond to the minimum of aerosols, this ensures that we have enough data within the file considered for the calibration while lowering the amount of aerosol contamination. Due to the presence of aerosols, we anticipate that the accuracy of this calibration procedure is low and the ocean backscatter coefficients are biased low. Assuming the average aerosol optical thickness of 0.13-0.14 [25] to be spread evenly within a 500 to 1000 m boundary layer and a value of lidar ratio of 20-25 sr [26, 27], the calibration is biased low by a factor 4 to 8 (i.e., up to one order of magnitude of calibration error). This aerosol background does not affect the bubble mask, as it does not impact the depolarization in the ocean, but this is important for discussing void fraction retrieval. The improvement of the void fraction retrieval accuracy is a crucial motivator to use the ocean surface as a calibration target, and the development of a correction accounting for the slope of the waves in the future.

## 6. Results and discussion

### 6.1 Lidar depolarization and multiple scattering considerations

As expected from [16], bubble clouds significantly affect the cross-polarization channel. Moreover, it is much more noticeable than in the co-polarization channel. This difference between co-polarized and cross-polarized channels is interesting, considering the signal intensity relates to the void fraction [28], and the bubble clouds were visible in the co-polarized channel in the laboratory environment [29]. For the dataset presented in this study, the co-polarization signal is slightly larger in presence of bubbles. As shown in section 6.5, it makes it possible to remove the contribution from the water molecules and biological content, to create a void fraction retrieval algorithm. It would however, be challenging to accurately determine the presence of bubbles from the co-polarization signal alone. This is a major difference with the results of [29], using the same system in the laboratory, three months prior to this deployment. We attribute this difference to the larger contribution of the biological content in the ocean to the co-polarized signal background, but it could also be that the void fraction of the bubble cloud in the laboratory environment was significantly larger than anything we observed in the Ocean.

The detection of depolarization of spherical particles in the backscatter direction implies multiple scattering of the lidar beam in an optically dense medium. For the NRL shipboard lidar, because the system is so close to the target, it is not detecting light scattered back at 180 degrees (i.e., the backscatter direction). The exact angle will depend on the distance to the target. How this slight angle deviation affects the measured signal can be shown by looking at the M12 element of the Mueller matrix of spherical particles [30]. As shown in Fig. 5, this element has a significant variability between 180 and 177 degrees. However, this single scattering calculation has little meaning if the multiple scattering regime applies to the lidar observations, as it will change the scattering geometry. However, it means that the NRL shipboard lidar should, within its sensitivity limits, be able to detect depolarization by bubble features optically thin enough to fall into the single scattering regime. In that case, observing the same bubble cloud with a change of the angle of observation due to the boat attitude or wave height change will provide information on the particle sizes.

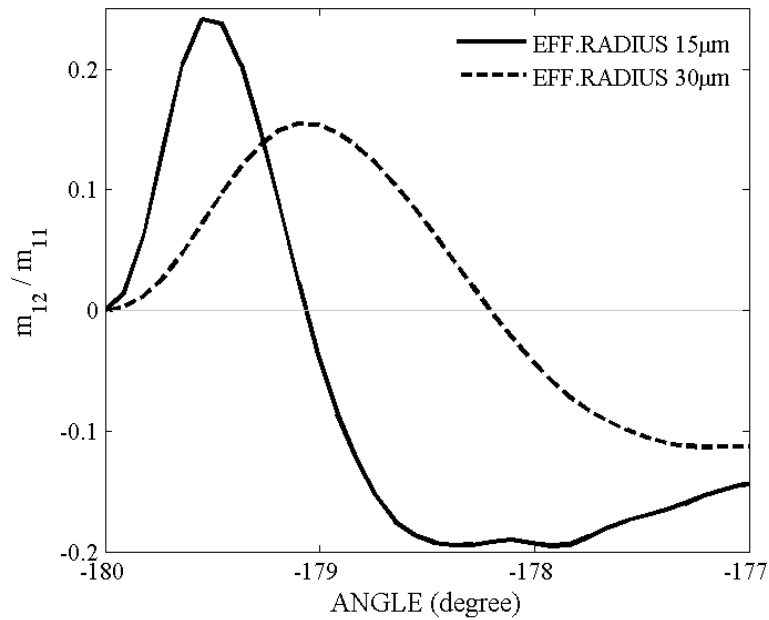

Fig. 5. Illustration of the depolarization element of the Mueller matrix for spherical particles near the backscatter direction. The calculation uses the distributions proposed by [30] (15 µm solid line, 30 µm dashed line).

We expect most instances of bubbles in the ocean to be optically dense and the single scattering considerations to not be relevant. In that case, an important matter to understand is which regime of multiple scattering lidar observations fall into [31, 32]. Because of the ambiguity of the scattering event's time, the lidar's bubble depth estimate become inaccurate in the presence of wide-angle multiple scattering. This is due to side scattering events being measured as if they are coming from a greater distance.

Wide angle multiple scattering of light occurs when the width of the "footprint" (X) projected by the field of view of the receiver at the range of the target is in the same order as the transport mean free path (MFP) of light [32]. In other words, the lidar can accurately measure the vertical extent of the bubble field only if

$$X \ll \frac{MFP}{(1-\omega_0 g)} \qquad (1)$$

The single scattering albedo $\omega_0$ for air bubbles is approximately equal to 1 [28, 30]. The asymmetry factor g is approximately 0.8443 [30]. The NRL shipboard lidar receiver has an angle of 140 mrad, corresponding to a telescope footprint of around 1.2 m. This value varies slightly with the waves' height, the lidar's attitude (pitch, roll, yaw), and the boat's heave. For the conditions of Eq. 1 to apply, the extinction due to the scattering of the bubble cloud must be much lower than 5.35 m$^{-1}$. The statistical characteristics of the profile of the lidar backscatter coefficient provides the order of magnitude of the extinction coefficient due to the bubbles, which can then be compared to this value. The statistic of the profile is shown in Fig. 6. Even if the backscatter itself decreases as a function of depth as fewer and fewer bubbles reach these levels, the profile is monotonic enough that the average slope provides the proper order of magnitude for this coefficient. Even if the signal in the bubble stays valid down to 20-30 m (see sections 6.2 and 6.3), it is clear from Fig. 6 that the best quality of the backscatter signal for clear water is limited to a depth of 5-10 m. No geophysical explanation exists for the change of backscatter coefficient slope around 5 m and 12 m. The increase of the backscatter coefficient as a function of depth (below 12 m) is a signal artifact. It is important to be aware that, even if this is typically not a limiting factor, the NRL shipboard lidar is designed to study features with stronger scattering than clear water.

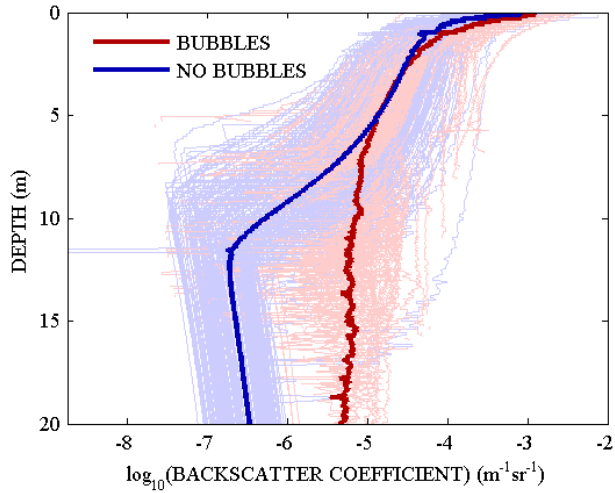

Fig. 6. Illustration of the average backscatter coefficient of bubbles and bubbleless lidar profiles.

The average slope of the logarithm of the backscatter coefficient is -0.24 m$^{-1}$, which corresponds to an extinction coefficient of 0.24 m$^{-1}$ if we assume a multiple scattering coefficient of 0.5 [31, 33]. This extinction coefficient is related to an MFP significantly below the threshold to create wide-angle multiple scattering even for a different value of the multiple scattering coefficient, and with a different assumption relating the scattering MFP to the lidar observations. This statement comes from the fact that the theoretical maximum of the multiple scattering coefficient should be close to 0.1 for full isotropization of light polarization in a dense medium [34, 35]. As a side note, the 90 m footprint of the CALIPSO lidar implies that this system is well within the wide-angle scattering regime for underwater bubble clouds.

Concerning small angle scattering, following [32], the criteria is

$$MFP \frac{\lambda}{\pi a} \lesssim X \qquad (2)$$

For the MFP corresponding to our observations, all bubbles with a radius > 0.5 µm are in the small angle scattering regime. Although the exact number of the such small bubbles is typically not measured by acoustic sensors, a peak of the bubble size distribution around 10 to 30 µm has been suggested in the past [30, 36]. The tank experiment we conducted in the breaking wave tank of the littoral high bay of the Laboratory of Autonomous System Research [29] seemed to peak around 1 - 2 µm based on acoustic resonance estimates, so even these artificially generated wave conditions would fall under this scattering regime.

In addition to the capability to measure the depth of bubble clouds, the impact of small-angle multiple scattering is an apparent reduction of the laser attenuation. Once the attenuation is corrected, the signal is identical to the single scattering return, and the underlying assumption of the formalism of [28] applies to our dataset (see section 6.5).

**6.2 Bubble feature detection**
Whitecaps and bubbles have a clear depolarization signature. This first version of the bubble mask relies on a simple threshold of depolarization as defined by the ratio of the cross-polarization channel to the co-polarization channel. After removing some detection artifacts (data with linear depolarization larger than 1), only the data with a depolarization ratio larger than 0.015 are considered bubbles. To determine the bubble depth, the algorithm also includes a criterion of continuity. The bubble depth corresponds to the number of continuous data points below the water surface whose depolarization value is above this threshold. This continuity criterion removes some false positives, which manifest as isolated points due to the noise. However, it prevents the detection of bubble clouds that do not connect to the surface. The results of the bubble mask are illustrated in Fig. 7. As previously mentioned, the intensity (co-polarization) and depolarization features are different, with no clear features visible in the lidar intensity. There are a few false positives remaining above the ocean surface, marked by the yellow/red curved on the intensity (Fig. 7a). The bubble depth calculation removes these false positives as they are not a continuous structure below the ocean surface. The conical structures of bubbles measured by the lidar from depolarization are qualitatively similar to previous studies of bubbles created by breaking waves [37, 38].

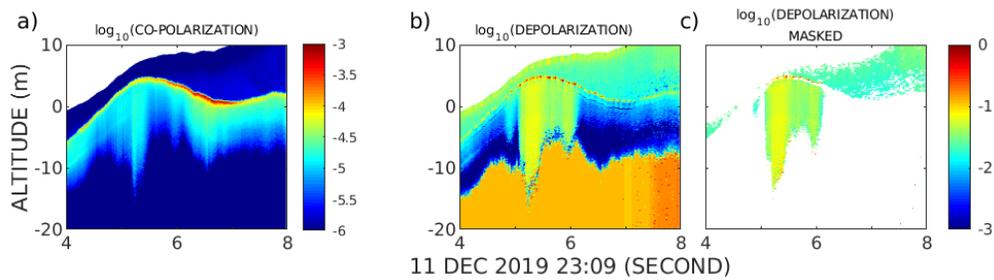

Fig. 7. (a): lidar attenuated backscatter intensity (decimal logarithm) for a segment starting at 11$^{th}$ Dec 2019 23:09Z, (b): same as a) but for the lidar depolarization, (c): same as b) but the non-underwater bubble features have been removed with the bubble mask (above water features are filtered at a later step). Both b) and c) share the same color code (decimal logarithm of the depolarization).

It is important to stress that the lidar has more capabilities than the simple approach presented here. As shown in Fig. 8, we can separate the continuum of observations into four broad domains: low surface and subsurface depolarization (bubbleless ocean), high surface depolarization with low subsurface depolarization (whitecaps), high surface and subsurface depolarization (extended bubble clouds) and low surface but high subsurface depolarization (underwater bubble clouds). The distinction between these different domains may be necessary for future research based on a more complex bubble mask, which would compare the lidar measurement with previous studies of whitecaps. Very few studies rely on lidar systems to study bubble properties, and the sensors used in most of the other research have different limitations. Typically, non-lidar instruments above the water surface can only detect the whitecaps with limited vertical sensitivity, but contrary to instruments deployed underwater, they have to advantage to not affect the water flow and the bubble properties. Active acoustic instruments can measure the bubble vertical profile, but the signal is dominated by the resonant frequency. The comparison with the lidar is interesting, especially as an indication of the bubble size distribution [29]. Underwater cameras can measure the bubble size directly, as long as it belongs to a limited range of bubble sizes.

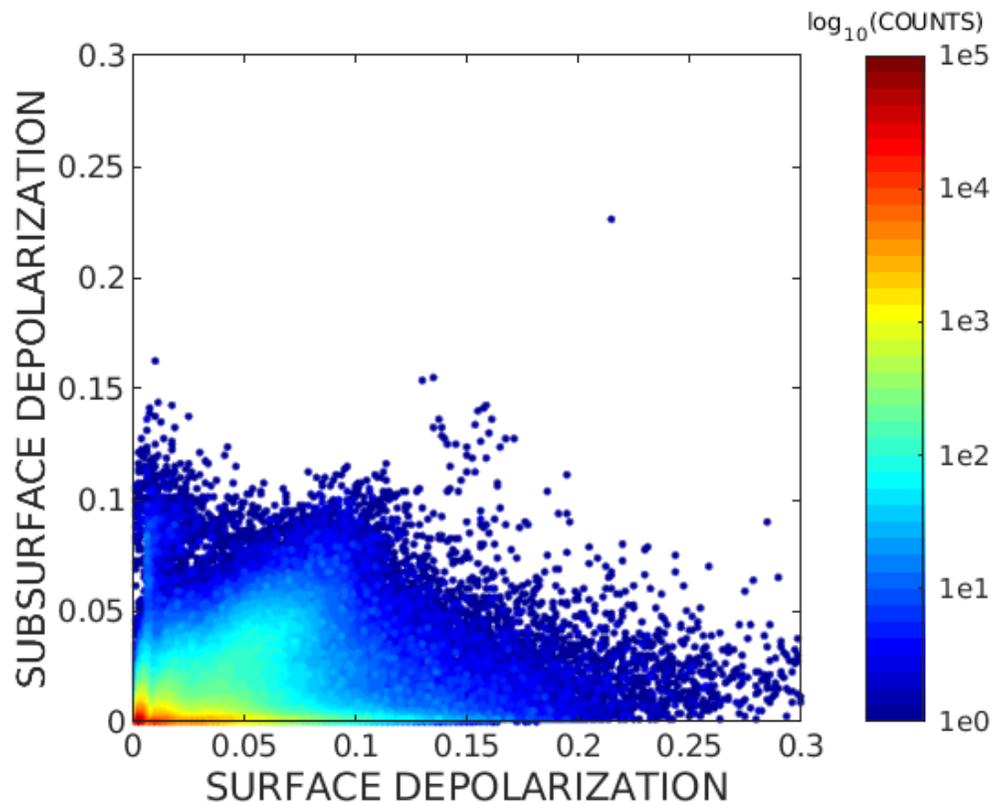

Fig. 8. Depolarization of the subsurface signal (higher for extended bubble clouds) as a function of the depolarization of the surface signal (higher for whitecaps). The color code is the decimal logarithm of the number of observations.

**6.3 Maximum penetration depth**
As described in the section 6.1, the NRL shipboard lidar observations fall into the regime of small-angle multiple scattering. Consequently, the lidar measures the vertical extension of the bubble field directly, and the presence of the scattering forward peak increases the penetration depth. The bubble feature detection algorithm is described in section 6.2. As previously explained, this bubble mask is a depolarization threshold as bubbles depolarize the lidar signal significantly. Fig. 9 illustrate one extreme case of bubble depth (Fig. 9a), and the bubble depth statistic from the bubble mask (9b). The bubble mask quantifies this system's maximum penetration depth in bubble clouds. As shown in Fig. 9a) and b), bubble depths over 25 m and up to 28 m are part of this dataset. However, most bubble cloud observations extend between 0 and 10 m, and the observed bubble depth does not reach 30 m. Because of the apparent low occurrences of these deep clouds and the novelty of the bubble mask, the data represented in Fig. 9b) do not allow to discriminate between a limitation of the lidar sensitivity, a physical limitation of bubble injection processes, or a limitation of the bubble depth in this specific dataset. This observed depth is extremely encouraging, considering that [39] maximum observed bubble depth from one year of echo sounders data very infrequently reaches 38 m deep, with most of their data also in the 0 to 10 m range. [1] present downward-looking echosounder data from the same deployment (R/V Sikuliaq) with a mean and maximum bubble plume penetration depths that exceed 10 m and 30 m. Note that even if there's a very good agreement with lidar and echo sounders data in bubble plume structure [29], the exact detection threshold of bubbles is instrument and algorithm dependent. The lidar algorithms and the dataset presented in this study are too new to exactly quantify the detection threshold in term of bubble properties, but it seems extremely close to the detection algorithms discussed in [1, 29, 39]. Ideally, we can refine the void fraction algorithm accuracy in the future to provide a void fraction threshold for the bubble mask.

The lack of correlation between bubble depth and wind speed in our dataset is unusual and has two origins. First, we were in conditions of a very steep swell, above the breaking threshold, and we had formation of bubble clouds in relatively low wind speed conditions. This is very interesting in term of ocean physics, but beyond the scope of to this paper, which focuses on what the lidar observations were. Second, if we consider a bubble cloud (Fig. 9a), there are many values of the bubble depth for a single bubble cloud. This paper presents high resolution observations, whereas the max bubble depth or average bubble depth is the quantity that is typically correlated with wind speed [39].

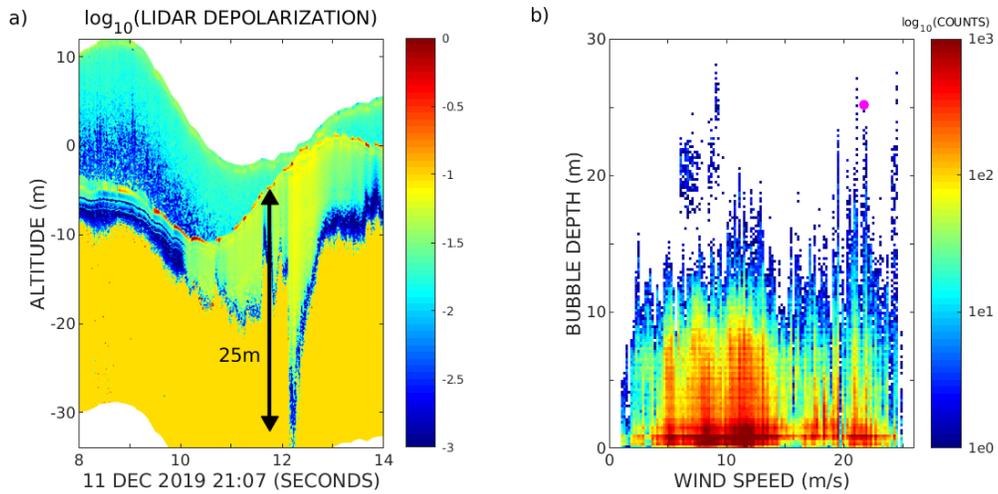

Fig. 9. a) Depolarization of the lidar signal for one extreme case of bubble depth. Green features are bubbles, and blue features the ocean (bubbleless) water. b) Statistic of bubble depth from the bubble mask (i.e., only features going from the surface to some depth) as a function of wind speed. The magenta dot in b) is the depth of the bubble cloud shown in a). The color code are a) the decimal logarithm of the depolarization and b) the decimal logarithm of the number of observations.

**6.4 Whitecaps contribution in the lidar equation**
Interestingly, no strong gradient of the surface return intensity is associated with the presence of the bubble clouds. Note that a moderate increase in the signal intensity is still present, which allows us to calculate a void fraction (see section 6.5). The signal intensity increase appears more clearly in an experiment that we conducted in the breaking wave tank of the NRL Laboratory for Autonomous System Research in September 2019 [29]. The smooth transition between the bubble and bubble-less ocean is a new result as the current lidar equation formalism [11, 15] creates a clear boundary between the specular reflectance of the ocean and the reflectance of whitecaps. It can be seen from Eq. 2 and Eq. 21 of [15] that the specular reflectance term will become close to 0 as W converges towards 1. If this formalism were correct, a clear horizontal intensity gradient would be associated with the strong depolarization induced by the bubbles. As we can see from Fig. 10, there is no correlation between the horizontal gradient of the ocean surface return and the gradient of surface depolarization

(correlation coefficient R is -0.1871, the coefficient of determination $R^2$ is 0.035), so a continuum of states would be more appropriate to describe the physics of bubbles in the ocean. This seem to imply that the correct formalism would be for the specular reflectance to not be a function of the presence of bubbles, and the whitecaps would continue to be an additive term. In other words, [15] derived the following equation to describe $\gamma_s$ the lidar specular reflectance, for a θ off-nadir angle

$$\gamma_s = \frac{(1-W)\rho}{4\cos^5\theta} p(\zeta_x, \zeta_y)$$

The fraction of the surface covered with whitecaps is $W$, and p(ςx, ςy) is the probability of slopes of waves ςx and ςy in both along- and cross-wind directions, respectively. ρ (sr⁻¹) is the Fresnel reflectance coefficient at nadir angle. Neglecting the subsurface return, the meaning of this equation is that for a given surface of the ocean, there are patches with whitecaps which have the whitecaps surface reflectance and patches without whitecaps which have the specular reflectance.

Assuming the whitecaps reflectance is correct, it suggests that the way the whitecaps coverage fraction $W$ appears in the specular reflectance is not appropriate, and at least for high resolution datasets, would be more accurately described with

$$\gamma_s = \frac{\rho}{4\cos^5\theta} p(\zeta_x, \zeta_y)$$

Note that it could also be an issue with the definition of $W$. It may be ill-suited for high resolution data as in that case, $W$ should be either 0 or 1.

Addressing the lidar equation will be the subject of further study. [15] did not address the validity of the whitecaps term, and we would need to revisit again the lidar equation to ensure consistency between the lidar and passive observation of whitecaps based on both theory and this dataset. The IMPACT project also included the collocated observations of the whitecap's coverage fraction from a camera with lidar bubble profile observations in 2022 in Greenland and Iceland, and the analysis of this dataset could be necessary for the next step of the work on the lidar equation. In the meantime, the fact that the bubble contribution is an additive term allows us to use [28] as is, as long as a surface correction is included. [28] mentions surface contamination as an error source, which is possible only if the specular reflectance and the bubble signal both contribute to the lidar signal. This underlying assumption is definitely supported by this dataset (Fig. 10), and it is enough to adapt his formalism to our dataset. However, the inconsistency with the lidar theory will need to be addressed in the future.

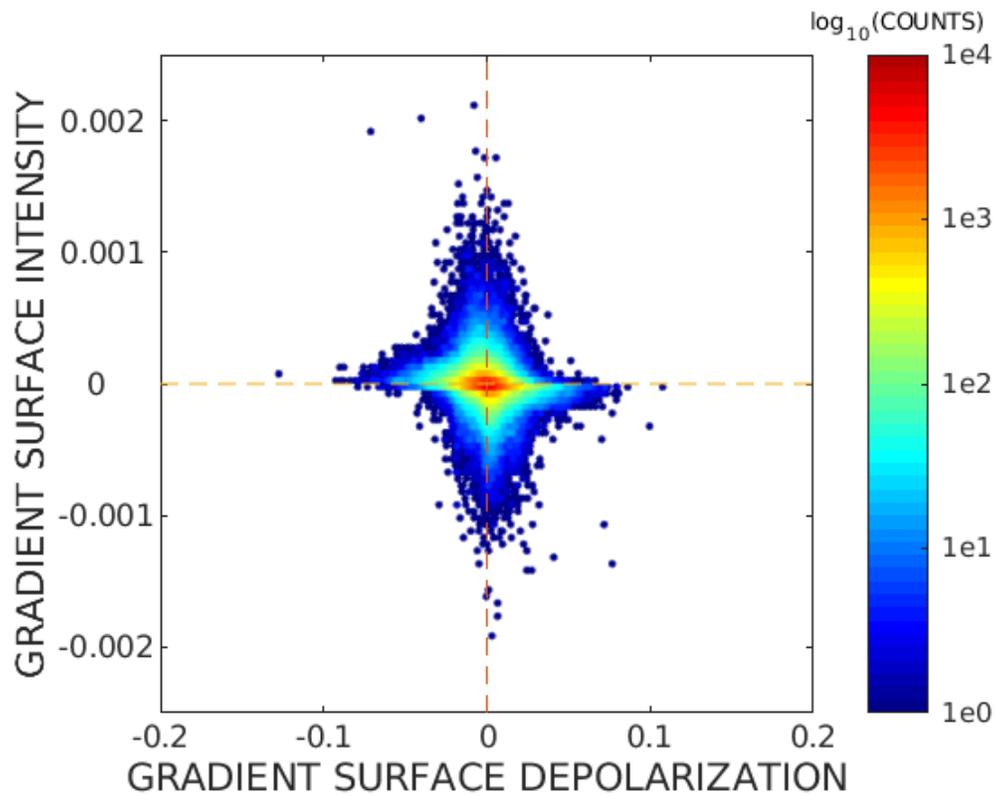

Fig. 10. Gradient of the surface intensity (co-polarization channel) as a function of the gradient of the surface depolarization. The color code is the decimal logarithm of the number of observations.

**6.5 Void fraction retrieval**
The relationship between the lidar backscatter coefficient and the bubble void fraction has been derived by [28]. The link is a simple multiplicative constant, and the lowest value of this constant is for bubbles without surfactants. Before this multiplicative constant can be applied to the NRL lidar observations, the signal has to be corrected from the total attenuation and the scattering contribution from the background (water molecules, biology) needs to be removed from the signal to keep only the signal from the bubble cloud.

In order to do so and for all profiles with bubbles:
- Using the bubble mask, the algorithm first selects all the profiles without bubbles (so this includes water molecules and biology). It then determines the extinction from the signal decrease as a function of depth from the average profile. This dataset's average extinction is around 0.1083 m$^{-1}$, consistent with the water chlorophyll content and the diffuse attenuation of previous studies [40].
- Using the backscatter intensity just below the bubble clouds, the algorithm then determines the attenuation of scattering by water molecules. Specifically, for each profile, we store the logarithm of the backscatter intensity 0.47 m below the lowest depth as determined by the bubble mask. Going slightly below the bubble cloud minimizes the likelihood of still having bubbles in the signal and allows to measure the backscatter of water molecules attenuated by the bubble cloud. Fig. 11 shows this attenuation value. The data suggests that there could be at least two or three regimes of attenuation of the bubble clouds (discontinuity of high attenuation around 4-8 m and low attenuation around 2 - 6 m). Due to the novelty of the bubble mask, it is preferable to not to reach too many conclusions at the moment. To simplify the correction procedure for this first version and to lower the likelihood of divergence of the retrieval, we use the retrieved extinction as low as 1.5 m. This depth corresponds to the highest number of data. In addition, only one attenuation regime seems to exist below this depth in Fig. 11. For deeper observations, the two-way extinction is set at 1.0892. This value corresponds to the decay of the logarithm average backscatter of bubbles between 2 and 5 m (Fig. 11). Furthermore, this lower attenuation value is consistent with less attenuation as the bubble density decreases. This simplification should be revisited in future versions of the algorithm.
- After correction of the total extinction, the average water molecules backscatter coefficient can be subtracted from the profile to retrieve the bubble backscatter coefficient and the associated void fraction.

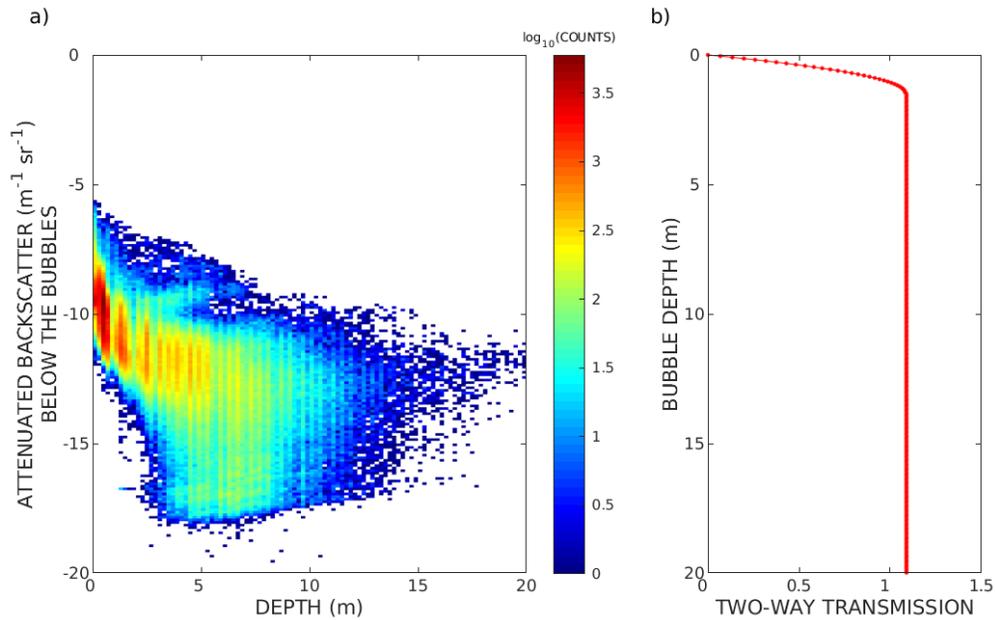

Fig. 11. a) decimal logarithm of the signal below the bubbles cloud. The color code is the decimal logarithm of the number of observations. b) Two-way transmission of the bubble cloud.

An example of the void fraction retrieval is shown in Fig. 12. for the data of Dec 11, 2019, at 21:13Z. In this case, the bubble cloud on the left (around second 17) has a much lower value of the void fraction at the surface than the bubble cloud observed a few seconds later (around second 19). This difference in void fraction could imply that it is older than the bubble cloud at the right of the picture.

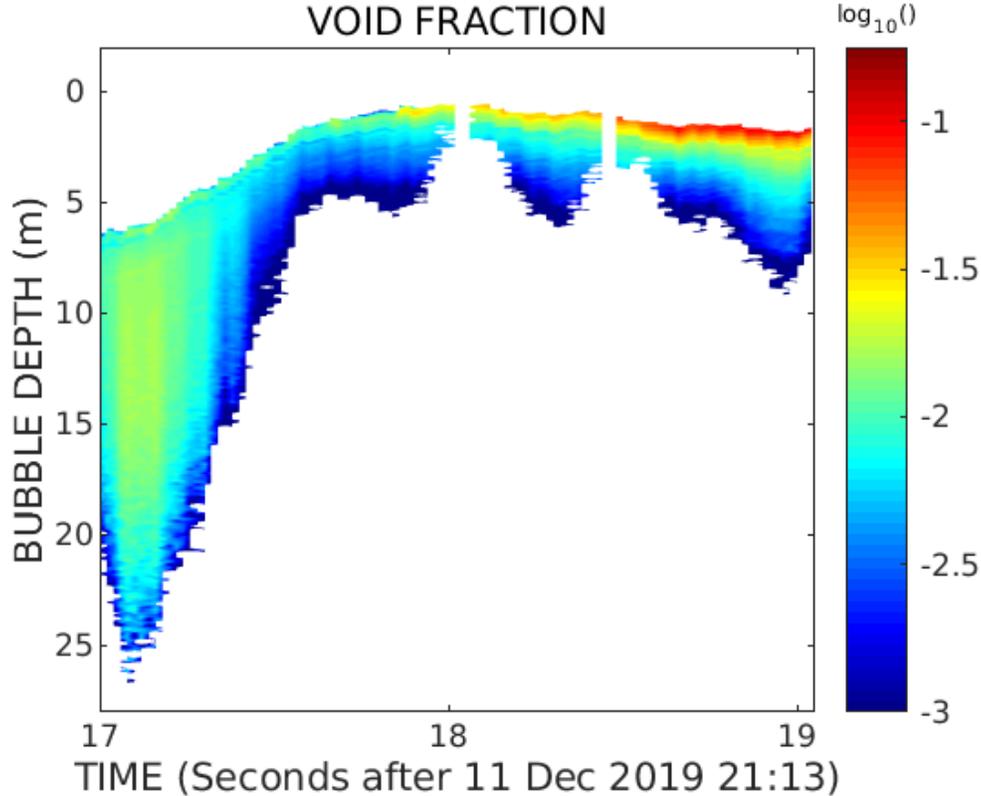

Fig. 12. Example of void fraction retrieval for a lidar case on Dec 11, 2019 (21:13:00Z). The color code is the decimal logarithm of the void fraction.

The void fraction estimates depend on several algorithms that are either new or newly applied to the NRL shipboard lidar. These algorithms include the bubble mask, the calibration procedure, and the correction for attenuation. We anticipate applying these algorithms to the whole shipboard lidar dataset (from other deployments, most without bubbles but with phytoplankton/zooplankton layers). The analysis of this extended dataset will provide further insight into the domain of validity of this algorithm and the associated uncertainty.

Beyond the current issue related to the lidar equation, there are also significant uncertainties due to the calibration coefficient and the lack of knowledge of the bubble surfactant. However, due to error compensation (i.e., the bias in these two factors partially compensates each other), we anticipate that the overall effect on the void fraction retrieval is a bias that cannot be higher than the calibration bias. As mentioned in section 5, the accuracy of our first algorithm is relatively low (factor 4 to 8 uncertainty of the calibration procedure, so almost one order of magnitude). The advantage of lidar for bubble research is that it is largely independent of the limitations of other instruments. There is no limitation concerning bubble properties related to depth or bubble size range. It is an advantage to currently available techniques to measure bubble properties. Lidar is limited in how deep they can monitor the water body. This limitation is a function of the hardware, software, and water turbidity. For bubbles, the strong scattering signal and the reduction of attenuation from the small-angle multiple scattering regime optimizes this feature's detectability. The ocean bubbles are an ideal target for an ocean lidar system. Lidar provide a new insight into the physics of the ocean at high wind speed, and the lack of lidar research in this topic (lack of dataset, lack of theory agreement with the observations) should guarantee fast progress.

## 7. Perspectives

As a short term, next step, this work allows us to determine the link between several bubble properties and the lidar measurements. Specifically, there is a link between bubble properties (bubble depth, void fraction) and the integrated lidar depolarization. This link creates estimates of bubble properties measured by a space lidar with depolarization. Preliminary results of global scale bubble depth maps are encouraging, and we anticipate presenting them in a future work.

Another interesting aspect of this study is the intrinsic difference between passive sensor observations of the bubble field in natural ocean conditions and what is detected by a lidar system that can penetrate the water surface and observe the whitecaps and various intensities of spray. Specifically, the water surface scattering properties that come from the statistic of the wave slope distribution do not exhibit drastic changes at the boundary between the bubbleless part of the ocean and waters with either whitecaps or extended bubble clouds. The ocean manifests a feature continuum above or below the water's surface from the lidar perspective. As previously discussed, this will help to guide future research related to the lidar equation.

## 8. Conclusions

Lidar is an ideal tool for obtaining information about the bubble environment. The bubbles create a strong, unambiguous depolarization, and the lidar simultaneously provides the context of the air-sea interface (surface height). A bubble mask is straightforward to create from a lidar with depolarization. It provides the vertical information of the bubble cloud structure that is consistent with other research based on acoustic echo sounders data. The multiple scattering regime of lidar observations can be derived based on both theory, and a statistic of the observations. The void fraction retrieval algorithm is complex, but does converge towards a solution that is relatively reasonable. It does not create negative values, or instabilities. The uncertainty is large but typical of backscatter lidar limitations (calibration accuracy, attenuation correction, scatterer refractive index). The relatively high uncertainty in this first version of the algorithm is due in large part to the novelty of the application. Subsequent versions will be more accurate. Bubbles have the advantage of showing a very strong scattering signature, and there are physical limits that bind the void fraction retrieval. A void fraction cannot be larger than 1, and the retrieval accuracy will increase as future lidar experiments validate our results. The smoothness of the observation is inconsistent with the currently derived lidar theory of whitecaps, and it suggests that future research should address this issue.

**Acknowledgements**
Funding of this research was provided by the NRL internal project (IMPACT). P.I. J. Thomson and co-P.I. Morteza Derakhti are greatly acknowledged for allowing the NRL researchers to be onboard, mount and use the NRL shipboard lidar on the bow of the R.V. Sikuliaq during the Wave breaking and bubble dynamics project (Breaking Bubble) funded by the National Science Foundation (NSF). We would like to acknowledge the help of Lucia Hosekova with the meta data. The crew of the RV Sikuliaq is greatly acknowledged for supporting this research. Victorija Morris is greatly acknowledged the creation of the 3D model of the mount of the NRL shipboard lidar.

**Disclosures**
The authors declare no conflicts of interest. The patent is pending for the algorithms presented in this study. It correspond to patent application # 18/476,959.

**Data Availability Statement**

The data gathered during NRL funded project are archived at NRL and are usually not available to the public due to the sensitive nature of NRL work. However, NRL has a process in place to request the data to be distributed (close or open release). If you have an interest in this dataset, please contact the main author damien.josset@nrlssc.navy.mil, with an explanation of the need, so that we can follow the appropriate release process.